\documentclass[preprint]{aastex}
\newcommand{\bvj}{B-V}

\newcommand{\bvt}{B_T-V_T}
\slugcomment{To appear in April 2006 {\it Astronomical Journal}}
\shorttitle{Erratum: Post-T Tauri Stars}
\shortauthors{Mamajek, Meyer, \& Liebert}
\begin{document}
\title{Erratum: Post-T Tauri Stars in the Nearest OB Association 
[Astron. J. {\bf 124}, 1670 (2002)]}
\author{Eric E. Mamajek}
\affil{Harvard-Smithsonian Center for Astrophysics}
\author{Michael R. Meyer, \& James Liebert}
\affil{Steward Observatory}
 
In Table 3 the F6V spectral standard star HR 7061 is mistakenly
aliased as HD 173677 but should be HD 173667.  In \S7.2 HIP 68282 is
mistakenly aliased as $\nu^1$ Cen (``nu 1'') but should be
$\upsilon^1$ Cen (``upsilon 1'') according to the Bright Star Catalog
(D. Hoffleit \& W.~H. Warren, Jr. [New Haven: Yale Univ. Obs.; 1991]).
The same error in SIMBAD has since been corrected.  In Appendix A,
third paragraph, the units of slope should be
K\,[log(cm\,s$^{-2}$)]$^{-1}$, not K\,[log(cm\,s$^{-1}$)]$^{-1}$.  In
Appendix C there is an incorrect sign in the second term in equation
C6, and incorrect color ranges were published for equations C6 and C7.
Equations C6 and C7 provide polynomial transformations between the
Tycho ($\bvt$) and the Cousins-Johnson ($\bvj$) colors.  The correct
equation C6, applicable over the color range ($\bvt$) $\in$ [0.40,
2.00], is
 
\vspace*{-0.5cm}
\begin{equation}
\bv = (\bvt) - 7.813\times10^{-3}(\bvt) - 1.489\times10^{-1}(\bvt)^2 + 3.384\times10^{-2}(\bvt)^3
\label{tycho2}
\end{equation}
 
\noindent The correct formula C6 was applied in the paper but
transcribed incorrectly to the manuscript.  Equation C7 was published
correctly, but the published color range is incorrect. For the color
range ($\bvt$) $\in$ [--0.25, 0.40], equation C7 should be used (same
as published):
 
\vspace*{-0.5cm}
\begin{equation}
\bv = (\bvt) - 0.006 - 1.069\times10^{-1}(\bvt) + 1.459\times10^{-1}(\bvt)^2
\label{tycho2}
\end{equation}

We thank John Carpenter for bringing the error in equation (C6) to our
attention, and Fran\c{c}ois Bonnarel and Christian Nitschelm for
pointing out the stellar misidentifications.
\end{document}